\documentclass[twoside]{article}
\usepackage{fleqn,espcrc2,psfig}

\title{Dynamical fermions on anisotropic lattices
\thanks{Talk presented by T.~Umeda.}}

\author{CP-PACS Collaboration : 
T.~Umeda\rlap,\address{Center for Computational Physics, 
   University of Tsukuba, Tsukuba, Ibaraki 305-8577, Japan}
S.~Aoki\rlap,\address{Institute of Physics, University of
   Tsukuba, Tsukuba, Ibaraki 305-8571, Japan}
M.~Fukugita\rlap,\address{Institute for Cosmic Ray Research,
   University of Tokyo, Kashiwa 277-8582, Japan}
K-I.~Ishikawa\rlap,$^{\rm a}$
N.~Ishizuka\rlap,$^{\rm a,b}$
Y.~Iwasaki\rlap,$^{\rm a,b}$
K.~Kanaya\rlap,$^{\rm b}$
Y.~Kuramashi\rlap,\address{High Energy Accelerator Research 
   Organization(KEK), Tsukuba, Ibaraki 305-0801, Japan}
V.~Lesk\rlap,$^{\rm a}$
Y.~Namekawa\rlap,$^{\rm b}$
M.~Okawa\rlap,\address{Department of Physics, Hiroshima University, 
   Higashi-Hiroshima,Hiroshima, 739-8526, Japan}
Y.~Taniguchi\rlap,$^{\rm b}$
A.~Ukawa\rlap,$^{\rm a,b}$ and
T.~Yoshi\'e$^{\rm a,b}$ }

\begin{document}

\begin{abstract}
  We report on our study of two-flavor full QCD on anisotropic
  lattices using $O(a)$-improved Wilson quarks coupled with an
  RG-improved glue. 
  The bare gauge and quark anisotropies corresponding to the renormalized 
  anisotropy $\xi=a_s/a_t = 2$ are determined as functions of 
  $\beta$ and $\kappa$, using the Wilson loop and 
  the meson dispersion relation at several lattice cutoffs and 
  quark masses.

\end{abstract}

\maketitle

\section{Introduction}
Study of quark gluon plasma using finite-temperature lattice 
QCD has seen a remarkable progress over the last decade. 
The equation of state (EOS) is one of the most studied topics, 
for which many calculations have been reported and have made clear 
the effects of dynamical quarks \cite{lat00}.
However, most studies on EOS in full QCD are made for the temporal lattice 
size of $N_t=4$ or $6$, and 
a reliable continuum extrapolation of EOS 
is prevented by large scale violation present for small $N_t$.

Recently, the CP-PACS Collaboration proposed the use of anisotropic
lattice for calculations of EOS.
This proposal was tested for the pure gluon system \cite{namekawa01}, 
for which a well-controlled continuum extrapolation of EOS 
was performed for the first time.

We wish to extend the use of anisotropic 
lattice to calculations of EOS for full QCD. 
This requires a tuning of the bare gauge and quark anisotropy parameters 
which corresponds to a given value of the anisotropy. 
In this paper we report results of such a tuning.

\section{Lattice action}

We adopt an O(a)-improved Wilson quark action coupled with 
an RG-improved action for gluons. 
This combination of actions has been adopted in a series of systematic 
investigations at $T=0$ \cite{cppacs02}.
It also shows better scaling properties in finite-temperature QCD,  
{\it e.g.}, the expected O(4) scaling is observed around $T_c$ 
\cite{cppacs00}.

Here, we extend the study to anisotropic lattices.
We focus on the anisotropy $\xi=a_s/a_t=2$,
where $a_s$ and $a_t$ are the spatial and temporal lattice spacings, 
respectively.
This choice of $\xi$ is based on a study of efficiency in calculations
of EOS as described in our previous study in quenched QCD \cite{namekawa01}.

We consider the following gauge action which includes plaquette, 
$P_{\mu\nu}(x)$, and rectangle loop, $R_{\mu\nu}(x)$: 
\begin{eqnarray}
S_G=& & \nonumber\\
& &\hspace{-13mm}
  \beta\{\gamma_G^{-1}\sum_{x,i>j}
             (c_0^s P_{ij}(x)
             +c_1^s \{R_{ij}(x)
                     +R_{ji}(x)\})\nonumber\\
& &\hspace{-13mm}
  +\gamma_G\sum_{x,k}
             (c_0^t P_{k4}(x)
             +c_1^t R_{k4}(x)
             +c_2^t R_{4k}(x))\},
\end{eqnarray}
where $c_0^s+8c_1^s=1$ and $c_0^t+4(c_1^t+c_2^t)=1$.
Carrying out the RG improvement program of Ref.~\cite{Iwasaki} 
on this action, we find that a simple choice $c_1^s=c_1^t=c_2^t=-0.331$, 
which is the same as for the isotropic case, 
sufficiently improves the theory for small anisotropy 
$\xi\approx 1 - 4 $ \cite{ejiri02}. 

The quark action we study is as follows \cite{harada01}.
\begin{equation}
  S_F = \sum_{x,y} \bar{q}(x) K(x,y) q(y),
\end{equation}
\vspace{-5mm}
\begin{eqnarray}
& & \hspace*{-7mm} K(x,y) = \delta_{x,y} - \kappa_t 
\{ (1-\gamma_4)U_4(x)\delta_{x+\hat{4},y} \nonumber\\
    & & \hspace{-5mm}
+(1+\gamma_4)U^\dagger_4(x-\hat{4})\delta_{x-\hat{4},y} \}
 \nonumber \\
    & & \hspace{-5mm}
     - \kappa_s \sum_{i} 
 \{ (r-\gamma_i)U_i(x)\delta_{x+\hat{i},y}\nonumber\\
    & & \hspace{-5mm}
 +(r+\gamma_i)U^\dagger_i(x-\hat{i})\delta_{x-\hat{i},y} \}
 \nonumber \\ 
   & &  \hspace{-7mm}
     -( \kappa_s c_t \sum_i \sigma_{4i}F_{4i}(x)
               + r \kappa_s c_s \sum_{i>j}
                   \sigma_{ij}F_{ij}(x)) \delta_{x,y}.
\end{eqnarray}
The bare anisotropy  $\gamma_F$ of the fermion field is defined 
by the ratio of the spatial and the temporal hopping parameters,
$\kappa_t = \gamma_F \kappa_s$.
We set the Wilson parameter to be $r=1/\xi$ \cite{harada01}.
The hopping parameters are related to the bare quark mass as
$\kappa_s = 1/[2(m_q + \gamma_F + 3r)]$.
We also define $\kappa$ by 
$1/\kappa = 1/\kappa_s - 2(\gamma_F+3r-4)$. 
This parameter plays the same role as $\kappa$ on the isotropic
lattice.
For free fermion, equating the rest and the kinetic masses leads to 
$\gamma_F^{-1}=\xi^{-1}(1+m_q^2/3)$ when $r=1/\xi$ \cite{matsufuru01}.

For the field strength $F_{\mu\nu}$, we use the standard clover-leaf
definition.
At the tree level, the coefficients of the clover terms, $c_s$ and
$c_t$, are unity.
We incorporate the mean-field improvement, 
$U_i \rightarrow U_i/u_{0s}$ ($i=$1,2,3) and 
$U_4 \rightarrow U_4/u_{0t}$, 
with the spatial mean-field factor defined with the plaquette 
in 1-loop perturbation theory $u_{0s}=(1-1.154/\beta)^{1/4}$, 
and $u_{0t}=1$ for the temporal mean-field.

\section{Calibration procedure}

To realize a consistent anisotropy, the bare anisotropy parameters 
$\gamma_F$ and $\gamma_G$ for fermion and gauge fields have to 
be calibrated, such that 
the anisotropy $\xi_F$ calculated in the fermionic sector coincides 
with $\xi_G$ calculated in the gluonic sector.

We perform this calibration as follows. 
Calculating $(\xi_F,\xi_G)$ for some parameter sets of 
$(\gamma_F,\gamma_G)$ at a fixed $\kappa$ and $\beta$,
we make a fit of results with linear functions of $\gamma_F$ and $\gamma_G$, 
and locate the point satisfying $\xi_F=\xi_G=\xi$, where $\xi=2$ in our case. 
This point is denoted as $\gamma_F^*$ and $\gamma_G^*$,
which are functions of $\beta$ and $\kappa$.

For the determination of $\xi_G$, we adopt 
Klassen's method \cite{klassen98} of 
matching the Wilson loop ratios according to 
\begin{equation}
\frac{W_{ss}(x,y)}{W_{ss}(x+1,y)}=\frac{W_{st}(x,t)}{W_{st}(x+1,t)}
,~~~~t=\xi_Gy
\end{equation}
where $W_{ss}(x,y)$ and $W_{st}(x,t)$ are spatial-spatial and
spatial-temporal Wilson loop, respectively.

For $\xi_F$ we use the relativistic dispersion 
relation of mesons, and demand 
\begin{equation}
 E(\vec{p})^2=m^2+\frac{\vec{p}^2}{\xi_F^2}+O(\vec{p^4}),
\end{equation}
where $E$ and $m$ are in units of $a_t$, while the spatial 
momentum $\vec{p}$ is in units of $a_s$. The latter is defined with 
$\vec{p}=2\pi\vec{n}/L_s$,  $\vec{n} = (0,0,0),$ $(1,0,0)$ and its 
permutations.
We evaluate $\xi_F$ using either pseudoscalar or vector mesons.  
The results are referred to as $\gamma_F^*(PS)$ and $\gamma_F^*(V)$
(and also $\gamma_G^*(PS)$ and $\gamma_G^*(V)$) for the calibrated
bare anisotropies.

\section{Simulation details}

Simulations are made with the HMC algorithm for two flavors 
of degenerate quarks, 
using an even-odd preconditioned BiCGStab quark solver.
The calibration is performed at $\beta=1.8,1.9$ and $2.0$ on
a $8^3\times24, 8^3\times24$ and $10^3\times30$ lattice, respectively. 
Six values of $\kappa$ are used at each $\beta$ corresponding to 
$m_{PS}/m_{V}\simeq 0.6, 0.7, 0.8, 0.85, 0.9$ and $0.92$. 
Measurements are performed at every 5 trajectories 
up to 1000 -- 1700 trajectories after 
300 thermalization trajectiries. 
Errors are estimated by the jackknife method with bins of 
50 trajectories, and error propagation is used in the fitting to $\xi=2$.

If the lattice scale is estimated from the Sommer scale 
$r_0=0.5$fm, 
the typical lattice size in the spatial direction is about 2.0 fm (2.3 fm) 
at $\beta=1.9, 2.0$ ($\beta=1.8$) for $m_{PS}/m_{V}\simeq 0.8$.

\section{Calibration results}

\begin{figure}[tb]
\centerline{\psfig{figure=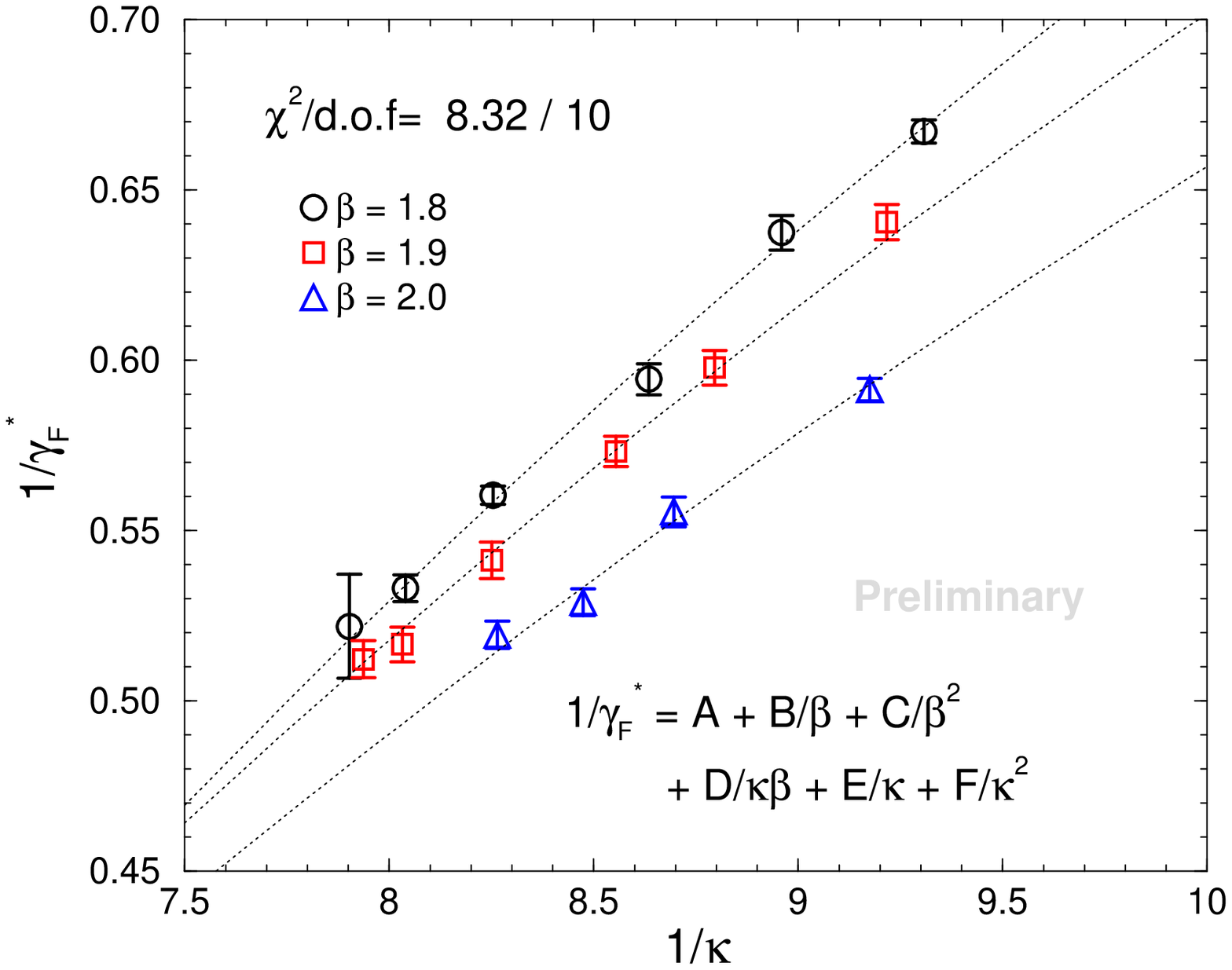,width=65mm}}
\centerline{\psfig{figure=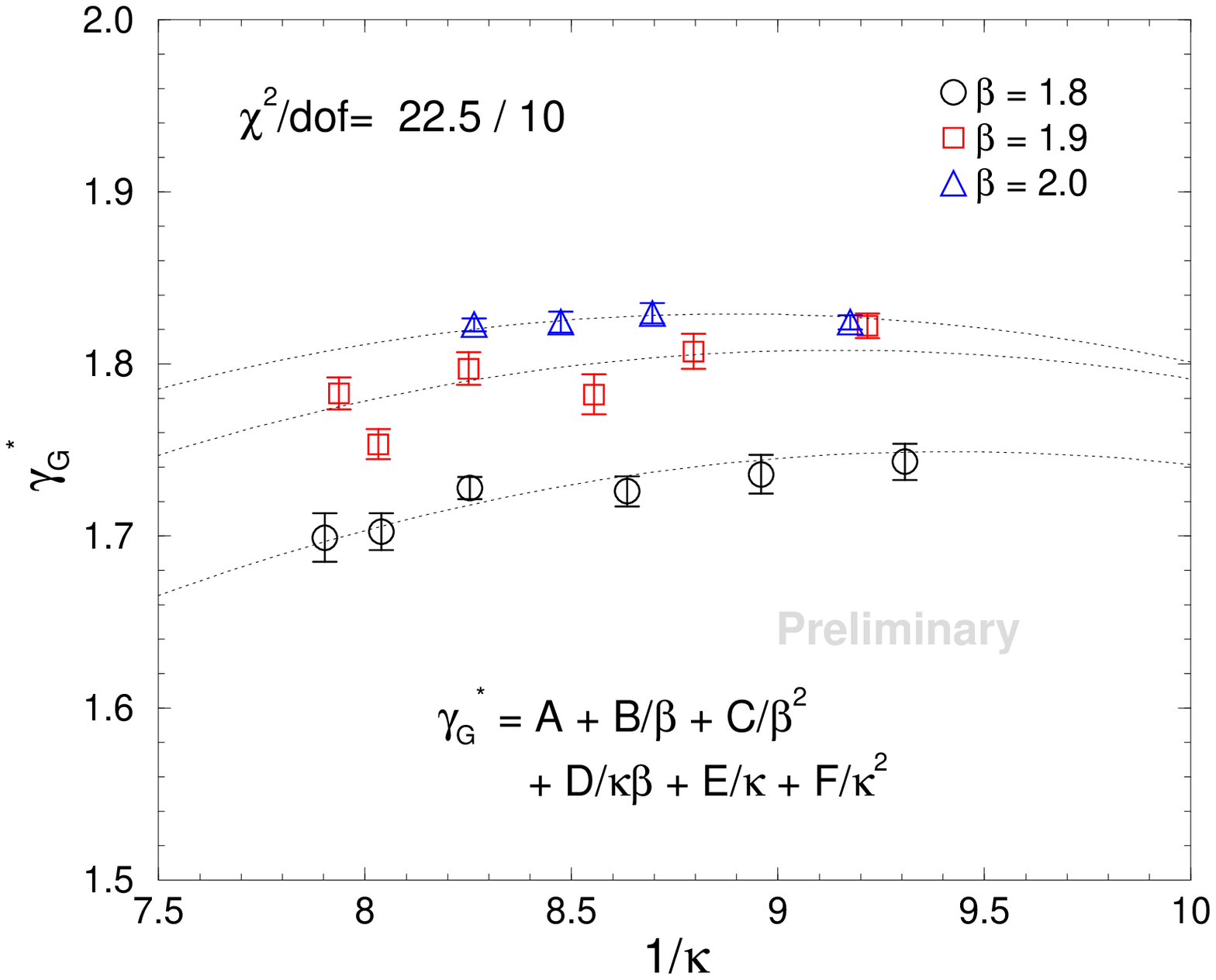,width=65mm}}
\vspace{-10mm}
\caption{$1/\kappa$ dependence of calibration results at each $\beta$}
\vspace{-5mm}
\end{figure}

Figure 1 shows the results for $\gamma_G^*$ and 
$1/\gamma_F^*$ as a function of $1/\kappa$ at each $\beta$. 
Unlike the case of quenched calculation \cite{matsufuru01}, 
where $1/\gamma_F^*$ shows no linear terms in $m_q$, 
linear terms are important in full QCD even with the choice $r=1/\xi$.
Therefore the results are fitted with a
general quadratic function of $1/\beta$ and $1/\kappa$,  
$f(\beta,\kappa)=A + B/\beta + C/\beta^2 + D/\beta \kappa 
+ E/\kappa + F/\kappa^2$.

\begin{figure}[tb]
\centerline{\psfig{figure=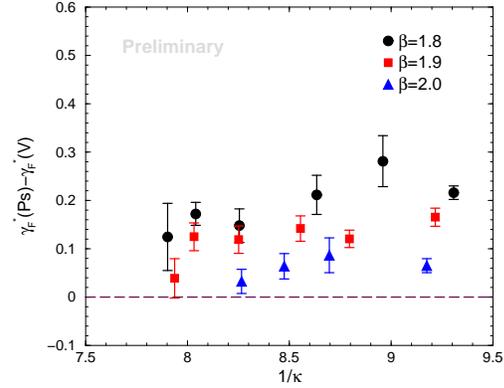,width=65mm}}
\vspace{-10mm}
\caption{Difference between pseudoscalar and vector in determination 
of $\xi_F$}
\vspace{-5mm}
\end{figure}

Figure 2 represents the difference between $\gamma_F^*(PS)$ 
and $\gamma_F^*(V)$.
Non-vanishing values of this quantity represnets an $O(\alpha a)$ lattice 
artifacts for our action choice. 
We confirm that the difference tends to zero 
( less than 5 \% for $\beta\geq 2.0$) as
$\beta$ is increased.
We plan to use this difference to estimate systematic errors 
in continuum extrapolations.

An extension of the present calibration to the region of weaker coupling 
(larger $\beta$) is under way.  

\vspace*{3mm}

This work is supported in part by the Large-Scale Numerical Simulation 
Project of the Science Information Processing Center (S.I.P.C) of 
University of Tsukuba
and by the Grants-in-Aid for Scientific Research by the ministry of 
Education (No.12304011, 12640253, 12740133,
13640259, 13640260, 13135204, 14046202, 11640294). 
The simulations are carried out on Fujitsu VPP5000 of S.I.P.C.


\begin{thebibliography}{99}

\bibitem{lat00}
 See e.g. S.~Ejiri,
 Nucl. Phys. B(Proc.Suppl.)94 (2001) 16.

\bibitem{namekawa01} 
 Y.~Namekawa et al. (CP-PACS Collaboration), 
  Phys. Rev. D64 (2001) 074507.

\bibitem{cppacs00}
 A.~Ali Khan et al. (CP-PACS Collaboration), 
 Phys. Rev. D63 (2000) 034502.

\bibitem{cppacs02}
 A.~Ali Khan et al. (CP-PACS Collaboration), 
 Phys. Rev. D65 (2002) 054505.

\bibitem{Iwasaki}
 Y. Iwasaki, Nucl. Phys. B258 (1985) 141; 
 Univ. of Tsukuba Report No. UTHEP-118 (1983).

\bibitem{ejiri02}
 S.~Ejiri, K.~Kanaya, Y.~Namekawa and T.~Umeda,
 in preparation.

\bibitem{harada01}
 J.~Harada et al.,
 Phys. Rev. D64 (2001) 074501.

\bibitem{klassen98}
 T.~R.~Klassen,
 Nucl. Phys. B533 (1998) 557.

\bibitem{matsufuru01}
 H.~Matsufuru, T.~Onogi and T.~Umeda
 Phys. Rev. D64 (2001) 114503.

\end{thebibliography}
\end{document}